\documentstyle[agupp,epsf]{article}

\lefthead{JONES}

\righthead{DEPLETION PATTERNS AND DUST EVOLUTION}

\received{February~1999}     
\revised{May~1999}
\accepted{June~1999}

\paperid{99.0069}

\cpright{AGU}{1996}

\authoraddr{A.P. Jones, Institut d'Astrophysique Spatiale, 
Universit\'e Paris XI, B\^at. 121, 91405 Orsay, France.  
(e-mail: ant@ias.fr)}
  
\slugcomment{To appear in the {\it Journal 
of Geophysical Research}, 1999.}

\makeatletter
\let\reset@font\empty
\makeatother

\begin{document}

\newcommand{\etal}   {{\it et al.\ }}
\newcommand{\etc}    {{\it etc.}}
\newcommand{\ie}     {i.\,e.}
\newcommand{\eg}     {e.\,g.}
\newcommand{\cf}     {{\it cf.\ }}
\newcommand{\araa}   {\itARA\&A}
\newcommand{\apss}   {\it Ap\&SS}
\newcommand{\ica}    {\it Icarus}
\newcommand{\um}     {\,$\mu\mbox{m}$}
\newcommand{\solmass}{\,$\mbox{M}_{\odot}$}
\newcommand{\kms}    {\,$\mbox{km\,s}^{-1}$}
\newcommand{\nod}    {\,$\mbox{cm}^{-3}$}

\title{Depletion patterns and dust evolution in the ISM}

\author{A.P. Jones}

\affil{Institut d'Astrophysique Spatiale, 
Universit\'e Paris XI, B\^at. 121, 91405 Orsay, France.}

\begin{abstract}
We review the use of elemental depletions in determining the
composition of interstellar dust and present a new interpretation of
the elemental depletion patterns for the dust forming elements in a
range of diffuse cloud types. We discuss this within the context of
dust processing in the ISM and show that Si and Mg are selectively
eroded from dust, with respect to Fe, as expected for a sputtering
erosion process. However, we find that Si is preferentially and
non-stoichiometrically eroded from dust with respect to Mg by some as
yet unidentified process that may act in conjunction with grain
sputtering. On this basis a new way of interpreting the depletions in
terms of `continuous' dust processing through erosion in the
interstellar medium is presented. The observed depletion patterns can
then be understood in terms of a gradually changing grain chemical
composition as the erosion of the atoms proceeds
non-stoichiometrically in the low-density interstellar medium. The
stoichiometric erosion of multicomponent (e.g., core/mantle) single
grains can qualitatively explain the observed depletions but is not
consistent with the preferential erosion of Si from dust. We present
suggestions for the usage of mineralogical terms within the context of
interstellar and circumstellar dust mineralogy.
\end{abstract}

\begin{article}

\section{Introduction}

Two connected quantities are pivotal in determining the mass and the
composition of dust in the interstellar medium (ISM). Firstly, the
absolute abundances of the elements with respect to hydrogen and,
secondly, the fractions of these elements that are incorporated into
dust. Many determinations of the dust composition based on
measurements of the gas phase abundances of the elements that are
normally found in dust have been made [e.g., Spitzer and Fitzpatrick
1993, 1995; Sofia, Cardelli and Savage 1994; Fitzpatrick 1995; Sembach
and Savage 1996]. However, in order to calculate the `missing mass',
or the fraction of an element that is actually incorporated into dust,
the absolute abundance of that element is required. A major problem
with this is that the absolute abundances of the elements in all
states (atomic, ionic, radical, molecular and solid state) along any
line of sight cannot currently be measured at the same time because of
the difficulty in observing the elements in the solid state
directly. There is however likely to be some progress made on this
problem very soon by the forthcoming Far-Ultraviolet Spectroscopic
Explorer (FUSE) and Chandra X-ray Observatory (AXAF) missions which
will be able to measure the abundance of an element in the gas and
solid phases, respectively.

The abundance of the elements in the Sun, combined with the elemental
composition of meteorites, provides an important `yardstick' for
interstellar (IS) dust studies because it is the best determined
reference system that we have, and one that encompasses most of the
elements in the periodic table [Anders and Grevesse 1989; Grevesse and
Noels 1993].  In Figure~1 and Table~1 we show the solar abundances of
Anders and Grevesse [1989] and Grevesse and Noels [1993] as a function
of atomic number (Z), and use these as a qualitative measure of the
chemical composition of the general ISM. The interesting `sawtooth'
pattern in Figure~1, peaked at even-numbered values of Z, or
equivalently peaked at multiples of approximately four atomic mass
units, arises from the `preferred' nucleosynthetic formation of
heavier elements from helium nuclei, $^4_2$He$^{++}$, during the
nuclear `burning' of hydrogen and helium within stars and
supernovae. The apparently anomalously high abundance of Fe is due to
its efficient formation in massive stars and supernovae.

Figure~1 allows some interesting general deductions to be drawn on the
likely composition of interstellar dust. C, H, O and N are abundant
and reactive elements and are the primary constituents of the detected
polyatomic IS and circumstellar (CS) gas phase species [e.g., Ohishi
1997]. Additionally, C and O are also primary constituents in
carbonaceous and silicate/oxide dust, respectively. N is not a major
dust component, possibly because it exists as N$_2$ and the strong
N$\equiv$N bond inhibits reactivity [Gail and Sedlmayr 1986]. Ne and
Ar are abundant but inert and do not participate in IS gas phase
chemistry, but they can passively accrete onto cold dust in dense
clouds. F, Cl and P are chemically reactive and principally found in
the gas phase. Sulfur can form solid sulphides, but in the ISM it does
not seem to be depleted into dust (see \S~2). The remainder of the
elements in Figure~1 form solid silicates and oxides, and are
constituents in well-known mineral phases. On the basis of the
abundances in Figure~1 we can divide the potential dust-forming
elements into four groups:
\begin{itemize}
\item primary dust constituents (abundance $\geq 300$ ppm) 
 C and O ({\it filled diamonds} in Figure~1).
\item major dust constituents (abundance $\sim 30$ ppm) 
 Mg, Si and Fe ({\it filled circles} in Figure~1).
\item minor dust constituents (abundance $\sim 3$ ppm) 
 Na, Al, Ca, and Ni ({\it heavy circles} in Figure~1).
\item trace dust constituents (abundance $\sim 0.1 -  0.3$ ppm) 
 K, Ti, Cr, Mn, and Co ({\it light circles} in Figure~1).
\end{itemize}
This simple classification is reflected in the IS silicate/oxide dust
composition determined from IS depletions studies [e.g., Spitzer and
Fitzpatrick 1993; Sofia, Cardelli and Savage 1994; Savage and Sembach
1996].

Solar abundances probably do not, however, apply to all phases of the
ISM. A recent attempt at deriving an absolute IS reference abundance
system was made by Snow and Witt [1996] (we refer to this reference
abundance system as `SW96'). Their approach was to assume that the
composition of the stellar atmospheres of newly-formed massive stars,
e.g., B stars [Kilian 1992, 1994; Adelman, Robinson and Wahlgren 1993;
Gies and Lambert 1992], completely sample the composition of the ISM
from which they formed. The derived abundances from this study are
about two thirds of the solar reference abundances of Anders and
Grevesse [1989] and Grevesse and Noels [1993], and are summarized in
Table~1. This result has major implications for IS dust models, most
of which have problems fullfilling these new abundance criteria [Snow
and Witt 1995]. This is the so-called `carbon crisis' which, in fact,
is not restricted to carbon but also applies to Mg, Si and Fe as
well. However, recent dust models by Mathis [1996] and Li and
Greenberg [1997] have been successful in overcoming this crisis and
meeting the new abundance criteria.

In the derivation of their reference abundance system Snow and Witt
[1996] assumed that there was no elemental fractionation during star
formation. This may not be an entirely valid assumption because of the
role of magnetic fields during cloud collapse and star
formation. Ciolek and Mouschovias [1996] determine that a relatively
large fraction ($\geq 90$\%) of the small IS dust grains may not reach
the cores of pre-stellar clouds, and are therefore not incorporated
into the forming star. The star will therefore be underabundant in the
heavy dust-forming elements, and the `cosmic' abundances determined
from the stellar photosphere will be underestimated.

\section{Depletions and Dust Composition}

The depletion of the dust-forming elements in a given region of the
ISM is determined by the dynamics of the accretion and erosion
processes acting on the dust in that region. In denser regions of the
ISM accretion is favoured by the increased collision rates between gas
phase species and grains. Erosion is more likely in the lower density
ISM due to the effects of shock waves. Thus, the depletion of the
elements in the ISM will be driven by these two competing processes.

Field [1974] showed that the elements with the highest condensation
temperatures generally have the highest depletion factors, and that
this is consistent with grain condensation in cool stellar
atmospheres. However, the observed effect is also consistent with the
lower temperature process of grain growth via accretion in molecular
clouds [e.g., Savage and Sembach 1996]. Thus, the elemental depletion
patterns observed in diffuse clouds could arise from the selective
accretion of elements with high condensation temperatures. The
accretion timescale for a cloud of density, $n_H$ (cm$^{-3}$), is
$\sim 10^9/n_H$~yr. For a diffuse cloud with a density of typically a
few tens of H atoms per cm$^{-3}$ the accretion timescale is therefore
of the order of $10^7-10^8$~yr, i.e., similar to the neutral ISM
cycling timescale of $3 \times 10^7$~yr which is driven by massive
star formation in molecular clouds [e.g., McKee 1989]. Thus, it is
likely that accretion onto grains in the very diffuse ISM or
intercloud medium could play only a minor role in determining
elemental depletions. It is also clear that for accretion to be
important some process must have previously transferred the elements
into the gas phase from the dust.

The chemical composition of dust can also evolve in the ISM as grains
lose atoms to the gas phase through the high energy processes that
occur in the supernova-generated shock waves that permeate the low
density ISM. They are particularly effective at processing dust in the
warm intercloud phase [McKee 1989]. Behind a shock front charged
grains are accelerated around the magnetic field lines as the
postshock gas cools and compresses. This leads to high energy
collisions between the gas atoms/ions and the grains resulting in the
sputtering (erosion) of the grain surfaces, and also to collisions
between grains which results in the vaporization and fragmentation of
grains [Jones et al. 1994; Jones, Tielens and Hollenbach 1996].
Fragmentation leads to significant changes in the dust size
distribution [Jones, Tielens and Hollenbach 1996] but it is sputtering
that can change the dust composition.  The effects of sputtering and
vaporization on the gas phase abundances of C, Mg, Si and Fe as a
function of shock velocity are shown in Figure~2 (data from Jones,
Tielens and Hollenbach [1996]). In Figure~2 the stoichiometric
sputtering of Mg, Si and Fe from the silicates was assumed. As the
shock velocity increases an increasing fraction of the dust forming
elements is transfered to the gas phase.

Evidence for the erosional processing of dust in the ISM comes from
the measured abundances of the dust-forming elements in the gas phase
as a function of the physical conditions along the line of sight
[e.g., Routly and Spitzer 1952; Cowie 1978; Crinklaw, Federman and
Joseph 1994]. These observations show increased gas phase abundances
of the dust-forming elements with increased cloud velocity and with
decreased density along the line of sight. In the latter case lower
density gas implies more warm intercloud medium along the line of
sight and therefore an increased probability of the dust and gas
having been exposed to supernova shock waves.

In Figure~3 we show the gas phase abundances of the major, and some
minor and trace, dust forming elements and sulfur for lines of sight
through cool clouds in the disk and warm clouds in the halo of the
Galaxy. The data in Figure~3 are taken from the extensive review of
Savage and Sembach [1996] but include updated Mg depletions using the
revised Mg$^+$ oscillator strengths of Fitzpatrick
[1997]. Interestingly, if one compares the data in Figures~2 and 3,
the warm cloud Fe depletions are consistent with dust that has been
shocked to $\sim 100$~km/s, whereas the Mg and Si data are consistent
with much higher shock velocities. This seems to be an indication that
Mg and Si are more easily liberated from dust than the stoichiometric
sputtering of silicates shown in Figure~2 indicates. The cool cloud
depletions are, on the other hand, representative of highly-depleted
unshocked dust. From the interpretation of the depletion data and the
inherent patterns many authors have shown that the IS dust composition
is consistent with a core and mantle grain structure [Spitzer and
Fitzpatrick 1993, 1995; Sofia, Cardelli and Savage 1994; Fitzpatrick
1995; Sembach and Savage 1996; Savage and Sembach 1996]. These authors
have shown that the depletion data is consistent with an Mg-rich
silicate mantle surrounding and Fe-rich silicate/oxide grain core. In
this work we have tried to avoid the use of the specific structural
labels for dust components, i.e., core and mantle. We prefer the use
of the generic terms `less refractory' and `more refractory'
components because these do not imply any structural relationship
between the different dust phases. Indeed core/mantle grain structures
probably do exist in IS dust but such structures are not unequivocally
indicated by the observed depletion data.

The data presented in Figure~3 are consistent with a more refractory
oxide/silicate phase and a less refractory silicate phase. The cool
cloud depletions give the maximal depletion silicate/oxide grain
composition, and the warm cloud depletions represent the composition
of a more refractory dust component that has survived erosion in the
ISM. As noted above this is consistent with the presence of shocked
dust in the halo clouds. The derived cool cloud and warm cloud dust
compositions are sensitive to the assumed elemental
abundances. However, the difference between the two sets of depletions
in Figure~3, the material eroded into the gas phase, represents the
composition of a less refractory silicate/oxide dust component. This
composition is not sensitive to the assumed abundances because its
composition is probed directly by the measured gas phase abundances,
assuming that a major fraction of the elements was originally in dust
[e.g., Spitzer and Fitzpatrick 1993; Sofia, Cardelli and Savage 1994;
Fitzpatrick 1995; Savage and Sembach 1996]. Table~2 summarizes the
elemental ratios for the silicate/oxide dust phase as a function of
the assumed reference abundance system.

Spitzer and Fitzpatrick [1993, 1995], Sofia, Cardelli and Savage
[1994], Fitzpatrick [1995] and Sembach and Savage [1996] have, for
example, shown that the derived elemental depletions can be used to
derive the composition, and also the structure, of IS dust. We
summarize the ideas used by these authors in the next paragraph, and
show how the conclusions depend upon the assumed reference abundances.

From the mineralogy of silicates and oxides the ratio R = (Mg+Fe)/Si
can be used to determine the interstellar grain composition. A value
of $\rm R = 2$ is indicative of an olivine-type silicate
stoichiometry, $\rm [Mg _x,Fe_{(1-x)}]_2 SiO_4$ ($\rm 0 \leq x
\leq 1$), whereas $\rm R= 1$ implies pyroxene-type silicate
stoichiometry, $\rm [Mg_x,Fe_{(1-x)}]SiO_3$ ($\rm 0 \leq x
\leq 1$). Values of R greater than 2 imply a mixed oxide and silicate
material, with the oxide content increasing with R. We present the
value of R and the Mg/Fe ratio for the data of Savage and Sembach
[1996], as a function of reference abundance, in Table 2 where we have
used the Mg$^+$ oscillator strengths of Fitzpatrick [1997] to derive
the Mg data. The overall dust composition (Table 2) is close to
olivine-type stoichiometry for the solar reference abundance, and a
mixed oxide/silicate material for the B star and SW96 reference
abundances. The most refractory dust component (Table 2) appears to be
an Fe-rich olivine-type silicate (solar reference), a mixed
silicate/oxide (B star reference) or an Fe oxide (SW96 reference). The
best determined dust phase, the least refractory material (Table 2),
is consistent with Mg-rich olivine-type stoichiometry (solar, B star
and SW96 reference abundances). We note that the dust compositions
that we derive here, i.e., Mg-rich and Fe-rich olivine-type silicates
and iron oxide, are essentially identical with those derived for the
young, intermediate mass pre-main sequence Herbig Ae/Be star HD~100546
[Malfait et al. 1998]. It is necessary to give a word of caution here
because although the depletion data may indicate a stoichiometry
consistent with a particular type of silicate (olivine or pyroxene),
this does not imply that the dust actually contains these specific
mineral phases. However, as noted, the similarity between the chemical
composition of the diffuse ISM dust, derived from depletions, and that
of the dust around the young star HD~100546, derived
spectroscopically, is a strong indication of the utility of depletion
studies in determining the mineralogical as well as the chemical
composition of dust in the ISM.

\section{Depletions and Dust Processing}

It is useful to compare the elemental depletion data for various lines
of sight through diffuse clouds for a range of dust forming elements
in order to search for systematics in the depletion patterns. To do
this it is desirable to normalize the data to the abundance of a
particular element [e.g., Si, Fitzpatrick 1995]. We prefer Fe for this
purpose because it is abundant in dust and is less readily eroded from
the dust than either Mg or Si, as we show later. In Figure~4 we show
the fractions of the elements Mg, Si, S, Ti, Cr, Mn, Fe and Ni in dust
normalized to that of Fe for the SW96 and solar reference abundances,
and for the four major diffuse cloud types defined by Savage and
Sembach [1996], namely; cool disk, warm disk, disk+halo and halo.  In
deriving Figure~4 the Savage and Sembach [1996] gas phase Mg
abundances were multiplied by a factor of two to allow for the revised
Mg$^+$ oscillator strength of Fitzpatrick [1997]. The cloud types
represent a range of IS cloud conditions with elemental depletions
decreasing from cool to halo clouds. It is clear from the depletion
patterns seen in Figure~4 that the overall trends are only weakly
dependent on the reference abundance system adopted. These same
patterns are also seen for the B star reference abundances. However,
in the B star and SW96 reference abundance cases there is clearly a
problem with the Si and Mg abundances. For these elements the maximum
observed gas phase Si and Mg abundances in the Savage and Sembach
[1996] data, allowing for the revised Mg$^+$ oscillator strength of
Fitzpatrick [1997], are greater than the relevant reference
abundances. In the case of Mg we do however note that the minimum dust
abundances in the Savage and Sembach [1996] data are lower limits, as
shown in Figure~4. Thus, the B star and SW96 reference abundances seem
to underestimate the Si, and possibly the Mg, abundances. In Figure~4
these effects are reflected in the negative fractions for Si and Mg in
the halo cloud data.  We note that the negative fractions simply
reflect uncertainties in the adopted reference abundances. We are here
only interested in the illustrated trends in Figure~4, and these are
essentially independent of the reference abundance system adopted.

The trends in Figure~4 are for the most part linear, except for Si and
perhaps Mg, indicating approximately constant fractional erosion of
the elements with respect to Fe. The Si depletion data shows a clear
`flattening' with decreasing depletion into dust.  A similar trend may
also apply to the Mg data because of the lower limit values for the
smallest depletions.  These data thus seem to indicate a saturation
effect in the erosion of Si (and Mg) from dust, i.e., Si (and Mg)
atoms at low concentration in a a Fe-rich matrix are hard to remove.
It is easier to understand this behaviour in the cases where the
reference abundance is less than solar because it clearly corresponds
to the removal of most of the Si (and Mg) from the dust --- for solar
abundances the saturation effect is more difficult to explain because
in this case about half of the Si still remains in the dust. This may
therefore be further evidence that the real IS abundances are less
than solar but that they are somewhat larger than the indicated B star
and SW96 values.  A close look at the SW96 data in Figure~4 shows that
there seems to be a preferential removal of Si atoms from dust with
respect to Mg. For example, for the cool disk cloud averages 87\% of
Mg and 90\% of Si are in dust. However, for the warm disk cloud
averages 52\% of Mg and 26\% of Si are in dust. This trend is also
true for the ranges of the observed values (the boxes in Figure~4).
Figure~4 also shows the results of a stoichiometrically eroded
core/mantle particle model. These show that such a model can match the
depletion patterns well, except in the case of Si. We therefore
conclude from these depletion patterns that Si is, initially at least,
more easily eroded from dust than Mg, and much more easily eroded from
dust than any of the other major, minor or trace elements in the
dust. In a study of depletions and the lifecycle of IS dust Tielens
[1998] reached a similar conclusion concerning the incorporation of Si
into a rather volatile dust component.

Grain erosion in the ISM is dominated by sputtering due to ion-grain
impacts in supernova-\-generated shock waves [Jones et al. 1994;
Jones, Tielens and Hollenbach 1996]. Thus, the changing elemental
depletions in the ISM should reflect the mechanics of this process in
which the lighter atoms are sputtered first (see for example the
results of the 20 keV proton irradiation of olivine in Bradley [1994]
which are consistent with this assumption). Thus, the elemental
sputtering sequence for the most abundant isotopes should be
\[
^{24}_{12}Mg > ^{28}_{14}Si \gg ^{48}_{22}Ti    > 
^{52}_{24}Cr > ^{55}_{25}Mn > ^{56}_{26}Fe    > ^{59}_{28}Ni
\]
but from the cool and warm disk cloud depletion data in Figure~4 the
elemental sputtering sequence seems to be
\[
^{28}_{14}Si > ^{24}_{12}Mg \gg ^{48}_{22}Ti \sim 
^{52}_{24}Cr > ^{55}_{25}Mn > ^{56}_{26}Fe \sim ^{59}_{28}Ni.
\]
This is as expected for the mechanical sputtering process, except for
the inversion of Si and Mg in the sequence. It therefore appears that
the erosion of Si from dust can not be completely explained by
sputtering but that some other process must also be operating. One
difference between Si and the other metals that might account for this
difference is that with silicate structures the Si atoms are always
bound to oxygen atoms, whereas the metals are always present as
interstitial cations with the structure. Thus, the effective
cross-section of an Si atom in an implantation interaction may be
larger. In other words the disruption of the silicate SiO$_4$
tetrahedra could facilitate the preferential removal of Si from the
structure leaving the interstitial Mg, Fe, etc. behind. One would
therefore expect to see this effect in the results of Bradley [1994]
if this were the case, but clearly this is not so. We therefore need
to invoke some other mechanism to explain the preferential loss of Si
from dust.

\section{The Silicon Erosion Problem}

The preferential erosion of Si from IS dust discussed in the previous
section is something of an enigma. Currently it appears difficult to
resolve this problem without further laboratory studies on the erosion
of silicates as a function of incident ion parameters. We suggest
several possible effects may that warrant further investigation:
\begin{itemize}
\item The abundances of Si and Mg in the ISM may be higher than in the
B star or SW96 reference systems. For example, 50\% enhancements would
bring the Si and Mg results more into line with those of Fe. This
effect could not however explain the preferential erosion of Si in the
disk clouds. Additionally, enhanced Si and Mg abundances do not seem
very likely.
\item The bulk of the interstellar Si and Mg, and some fraction of Fe,
exist a less refractory silicate, e.g.,
$\rm [Mg_{0.9},Fe_{0.1}]_2 SiO_4$, phase with most of the Fe in a
more refractory silicate/oxide phase [e.g., Sofia, Cardelli and Savage
1994; Fitzpatrick 1995; Sembach and Savage 1996; Savage and Sembach
1996]. The stoichiometric erosion of such a two component mix can
qualitatively explain the non-linear behaviour of Si and Mg (see
Figure~4). However, the Si is clearly eroded more efficiently that the
stoichiometric sputtering of an Mg-rich olivine can account for.
\item There is some chemically selective erosional process at work in
the diffuse ISM which results in the preferential disruption of the
silicate SiO$_4$ tetrahedra and the subsequent loss of Si atoms from
the grain.
\item Not all the Si is bound into silicates/oxides in the ISM. Some
fraction of the Si in an even less refractory material than Mg-rich
olivine could explain its faster removal from dust than Mg [e.g.,
Tielens 1998].
\end{itemize}

\section{Silicate Processing in the ISM}

Recent observations of silicates in emission in asymptotic giant
branch (AGB) star dust shells made with the Infrared Space Observatory
(ISO) [Waelkens et al. 1996; Waters et al. 1996: Malfait et al. 1998]
and in comets made with ISO, and with airborne and ground-based
observatories [Crovisier et al. 1996; Hanner, Lynch and Russel 1994]
indicate the presence of Mg-rich silicates that may be up to 50\%
crystalline. These observations raise some interesting questions
concerning the nature of IS silicates. For instance, in contrast to
the recent ISO results, observations of the silicate 10~$\mu$m Si---O
stretching and 18~$\mu$m O---Si---O bending modes in the ISM indicate
that IS silicates are amorphous [e.g., Mathis 1990]. Thus, some
process in the ISM amorphitizes crystalline silicates and/or some
process crystallizes amorphous silicates before their incorporation
into cometary bodies and the dust shells around young stars. It is
perhaps easy to understand how atomic and ionic collisions in shocks
and due to cosmic rays in the ISM could amorphitize a
material. However, it is difficult to find a process in which
amorphous grains could undergo crystallization without sustained
heating to temperatures greater than 1000~K [Hallenbeck, Nuth and
Daukantas 1998]. Thus, if the re-crystallization of amorphous IS dust
does occur it can only be in the vicinity of stars hot enough to heat
dust to high temperatures ($>$ 1000~K) for long periods, e.g., in the
compact H~II regions around hot, massive stars. The results of such
processing may already have been seen in the M17-SW [Jones et
al. 1999] and Orion [Cesarsky et al. 1999] H~II regions.

\subsection{Amorphitization of Dust in the ISM}

Clearly, some process in the ISM renders dust amorphous and alters its
chemical composition. This is certainly the case for silicate grains,
which have been shown to be partially crystalline in many CS shells
[e.g., Waelkens et al. 1996; Waters et al. 1996], but completely
amorphous silicate in the ISM [e.g., Mathis 1990]. However, whatever
the process that renders crystalline CS silicates amorphous once
injected into the ISM must also similarly affect the carbonaceous
grain species.  The effects of such processes on carbon grains have
not yet been fully investigated.

It therefore seems likely that dust in the ISM will retain no `memory'
of its formation in a given stellar environment, but may instead be
characterized by tracers of its processing in the intercloud medium or
low density IS clouds, i.e., implantation processing. The heavily
irradiationally processed morphologies seen in the glass with embedded
metal and sulphide (GEMS) interplanetary dust particle component that
have been proposed to be of IS origin [Bradley 1994] may directly
sample this processing. However, the IS origin for the GEMS processing
has yet to be proven. In contrast to the GEMS some presolar dust
components extracted from primitive meteorites, e.g., SiC and graphite
grains, show no such evidence of having been processed [Bernatowicz
1997].  These grains were formed around AGB stars, as their isotopic
compositions show, and have traversed the ISM without apparently
showing any evidence of processing despite their long lifetimes, $\sim
130$ to $\sim 2000$ million years [Lewis, Amari and Anders
1994]. Thus, the nature of the dust processing in the ISM is still
somewhat of an open question.

The most likely phenomenon for dust processing in the ISM is ion
implantation in IS shock waves, originating from supernovae and
stellar winds, and from cosmic rays. Table~3 outlines likely values
for some of the important parameters in this implantation processing.
This table shows that although the flux of cosmic rays may be lower
their energy and implantation depths are much greater than the
equivalent shock implantation parameters. Also, cosmic rays provide a
continuous flux of processing particles rather than the stochastic
interactions of shock waves. Thus cosmic rays will provide a source of
processing ions in all environments. Unfortunately, their flux at
relatively low energies ($\sim 1$ MeV) is not easily determined
because solar cosmic rays completely dominate at energies less than
100~MeV in the solar system [e.g., Moore 1999] and so it is very
difficult at this stage to quantitatively asses their effects on IS
dust. However, the fact that long-lived presolar grains show no signs
of processing (e.g., SiC and graphite particles) suggests that the
effects of cosmic rays are not important. These grains are then those
that were never exposed to the effects of shock waves in the ISM and
therefore arrived in the Solar System in a pristine state [Jones et
al. 1997].

\subsection{Dust Processing: A New Scenario}

The grain model of refractory Fe-rich oxide/silicate cores surrounded
by silicate mantles [e.g., Spitzer and Fitzpatrick 1993, 1995; Sofia,
Cardelli and Savage 1994; Fitzpatrick 1995; Sembach and Savage 1996;
Savage and Sembach 1996] is based on an interpretation of the observed
elemental depletions of, primarily, Mg, Si and Fe. This model is not
without its problems, for instance, in Mg silicate smoke condensation
experiments [Hallenbeck, Nuth and Daukantas 1998] the formed grains
show pure silica cores surrounded by Mg silicate mantles, which seems
to be inconsistent with the proposed IS core/mantle grain
structure. Alternatively, if the silicate mantles are formed by the
re-accretion of gas phase species onto oxide/silicate cores in the ISM
it is hard to see how pure silicates, uncontaminated by carbon and
other `non-silicate' elements, could form. It is also hard to see how
the stoichiometric erosion of a single or multiple component grain can
completely explain the observed depletions (e.g., \S~3 and
Figure~4). Given these apparent inconsistencies we suggest an
alternative model to explain the observations.

We propose that dust is formed in stellar sources with an overall
olivine-type stoichiometry consistent with `typical' IS Mg, Si and Fe
elemental abundances [e.g., Snow and Witt 1996]. We assume that the
dust is a mix of amorphous and crystalline olivine, pyroxene and oxide
grains as shown by the recent ISO observations [e.g., Malfait et
al. 1998]. These grains are then ejected into the ISM where they are
exposed to erosional processes in shock waves and in the low-density
gas which preferentially remove Si with respect to Mg, and both Si and
Mg with respect to Fe, principally through the effects of sputtering.
However, the enhanced removal of Si with respect to Mg is probably not
due to some purely mechanical sputtering process. The exact nature of
the Si fractionation process is unknown, but it is likely to be a
chemically-selective process or an indication that not all of the Si
is in refractory silicate/oxide phases. An increasing degree of
erosion and processing in the ISM (from cool disk to halo clouds)
therefore leads to grains with an increasing value of
$\rm R = (Mg+Fe)/Si$.
This is due to the preferential erosion of the least resistant
components first, e.g., the Mg-rich olivine-type silicates. In regions
where the dust is heavily processed the remnant grains may be composed
almost entirely of Fe oxides. They will contain only a minor fraction
of Si and Mg dispersed throughout the grains that is difficult to
remove because it is so dilute. There is therefore a progressive
enrichment in Fe with respect to Mg and Si as erosion proceeds. A
gradual evolution in the grain composition such as this is consistent
with the changes in the observed depletions on going from from
high-depletion, cool clouds to low-depletion, halo clouds [Savage and
Sembach 1996]. The observed `saturation' effect seen at low Si
depletions in the halo clouds seems to be consistent with an Si
abundance of the order of 2/3 of the solar value (Figure~4), i.e.,
greater than the SW96 reference abundance value of about 1/2 solar.

In this model the observed IS depletion patterns are therefore assumed
to arise from a progressive and continuous evolution of the overall
chemical composition of the grains in the low density ISM, and not
from the stoichiometric erosion of core/mantle dust particles. We
therefore do not need to invoke a core/mantle grain structure, but we
cannot rule out the possibility of non-stoichiometric erosion from
core/mantle grains. We are currently developing this model for the
progressive evolution of dust composition in conjunction with new data
from dedicated laboratory experiments.

\section{The Silicate Cycle in the ISM}

The cycle of silicate grains in the ISM can be summarized as follows:
\begin{itemize}
\item New dust is formed around `old' stars principally in their AGB
phase and a significant fraction of this dust may be formed in a
crystalline state. The overall dust composition is probably of
olivine-type silicate, reflecting the `typical' Mg, Si and Fe
abundance ratios. Oxides of Mg and Fe may also be formed
contemporaneously with the silicates in these CS environments.
\item This dust is ejected into the ISM in the later stellar evolutionary
stages through radiation pressure and stellar winds.
\item In the ISM the dust is subject to erosive processing (sputtering)
and implantation in super\-nova-\-generated shock waves and by cosmic
rays. The effects of these processes are to reduce the total grain
mass and to alter the overall grain composition (e.g., the
preferential removal of Si with respect to Mg, and of both of these
elements with respect to Fe). Additionally, the silicate grain size
distribution will be modified by fragmentation in shock waves.
\item ISM dust is incorporated into molecular clouds and then into new
stars and CS shells through cloud collapse and star formation. During
this phase the re-accretion of atoms eroded from the dust and other
gas phase species occurs in conjunction with grain coagulation.
\item Around young, hot, massive stars the dust is subjected to heating
to high temperatures in compact H~II regions and may undergo partial
crystallization.  Around the less massive stars near the end of their
lifetime (in the AGB phase) new dust is formed. This dust is then
eventually injected into the ISM, thus completing the cycle.
\end{itemize}

\section{Interstellar Silicate Mineralogy}

In view of the well-defined mineralogical terms now finding their way
into the astrophysical literature, principally due observations of the
dust shells around young and old O-rich stars made with ISO [e.g.,
Waelkens 1996; Waters 1996], it is perhaps useful to define specific
guidelines for the usage of mineralogical terms in the astrophysics
literature. For example, the olivine minerals,
$\rm [Mg_xFe_{(1-x)}]_2SiO_4$,
have very specific chemical compositions in the geological context,
e.g., forsterite
($\rm x = 0.9-1.0$) and fayalite ($\rm x = 0.0-0.1$).
The same holds true for pyroxene and almost all other mineral
phases. Unfortunately IS observations are often not detailed enough to
identify specific minerals by their chemical composition. This is
because the identifications are made through the use of infrared
spectroscopy, which often does not permit the exact determination of
the chemical composition, or via depletion studies which indicate
chemical composition but not mineralogical structure. We therefore
tentatively suggest that in the less than geologically exact science
of IS and CS mineralogy (astromineralogy?) that caution be exercised
in the naming of dust components and in the use of very specific
mineralogical names.

As an example of the silicate nomenclature that could be useful in
describing, specifically, the derived compositions of IS and CS dust
we propose the following broadly-based scheme:
\begin{itemize}
\item Where the grain composition is determined by comparison with
laboratory spectroscopic data on specific minerals, then the terms
Mg-rich, Fe-rich etc. be applied to the generic mineral name when it
is clear that the mineral is rich in a particular cation, e.g.,
Mg-rich olivine, Fe-rich pyroxene, etc. Otherwise just the generic
mineral name should be used, e.g., olivine or pyroxene.
\item Where the grain composition is determined from measurements of
depletions (no spectroscopic information available) and only an
inferred chemical composition can be derived which is dependent on the
adopted reference abundance. The dust should then be referred to as of
a particular mineral `-type,' and where relevant the cation enrichment
indicated, e.g., Mg-rich olivine-type silicate, olivine-type silicate,
Fe-rich pyroxene-type silicate, etc.
\end{itemize}
In the second case, based on depletion studies alone, it is almost
impossible to discern the difference between a real silicate and a mix
of oxides with the same stoichiometry, and indeed IS silicates might
easily be considered in terms of a mixture of oxides.

The adoption of a scheme such as this will hopefully avoid
`over-specific' labels being attached to IS dust compositions. It is
also hoped that it would reflect the unavoidably inexact nature of
cosmic mineralogy in comparison with the precisely defined mineralogy
on the Earth and in the Solar System.

\section{Connection to the Local ISM Dust}

Frisch et al. [1999] have used the standard depletion arguments to
interpret the nature and composition of the dust in the local ISM in
terms of the simple core/mantle grain model. They relate their
findings to the direct observation of IS dust in the Solar System with
the Galileo and Ulysses space missions. Table~4 shows their derived
abundances for the grain cores and mantles based on the SW96 reference
abundances. From the data presented in Table~4 the dust in the local
ISM seems to be primarily composed of metal oxides, with more than
60\% of the Si residing in the gas phase. In comparison with the above
discussions this implies that the local dust has been heavily
processed which is entirely consistent with the models for a shocked
local ISM discussed by Frisch et al. [1999].

As Frisch et al. [1999] point out, there is a good correlation between
the Mg$^+$ and Fe$^+$ in the local cloud which indicates that these
elements are eroded from dust in a constant ratio. However, this
constant erosion ratio does not apply to Si because of its high gas
phase abundance. Thus, the same behaviour seen in the Savage and
Sembach [1996] data as presented in Figure~4 is also manifest in the
local ISM.

\section{Conclusions}

We have reviewed the use of elemental depletion patterns in
determining the composition of interstellar dust in the diffuse ISM. A
new interpretation of the elemental depletions of the dust forming
elements in a range of diffuse cloud types is presented. We find that
Si is preferentially eroded from dust, with respect to Mg and that
both of these elements are preferentially eroded with respect to Fe,
as generally expected for a sputtering process. However, the enhanced
erosion of Si with respect to Mg is not yet understood and seems to
require some as yet unidentified process that may act in conjunction
with grain sputtering. It could also indicate that some fraction of
the Si is in a more volatile non silicate/oxide phase. We suggest that
further dedicated experimental work is necessary in order to
understand the details of the ion-grain implantation and erosion
interaction in the ISM.

On the basis of the analysis presented here a new way of interpreting
the depletion patterns in terms of `continuous' dust processing in the
low density interstellar medium is suggested. In this scheme the
processing is non-stoichiometric and leads to a gradual chemical
evolution of the overall silicate/oxide composition from olivine-type
silicate towards an Fe-rich mixed oxide phase. From the presented Si
depletion `saturation' effect we suggest that the Si abundance is of
the order of 2/3 of the solar standard, and not 1/2 as implied by the
SW96 Si reference abundance.

We discuss the likely origins of the processing of dust in the ISM and
conclude that the effects of cosmic rays may be negligible. This is
based on the fact that certain presolar grains extracted from
meteorites (e.g., SiC and graphite) show no sign of processing in the
ISM despite their long lifetimes. This then implies that
super\-nova-\-generated shock waves may be the principal agent for
dust processing in the ISM and that the presolar SiC and graphite
grains escaped the effects of fast shocks in the ISM.

Suggestions for the usage of mineralogical terms with\-in the context of
interstellar and circumstellar mineralogy are presented.


\setcounter{equation}{0}

\acknowledgments
The author wishes to thank the two anonymous referees of this paper
for their very helpful comments. APJ is grateful to the Soci\'et\'e de
Secours des Amis des Sciences for funding during the progress of this
work.


\end{article}

\newpage


\begin{figure}
  \centerline{\epsfxsize=14.0cm \epsffile{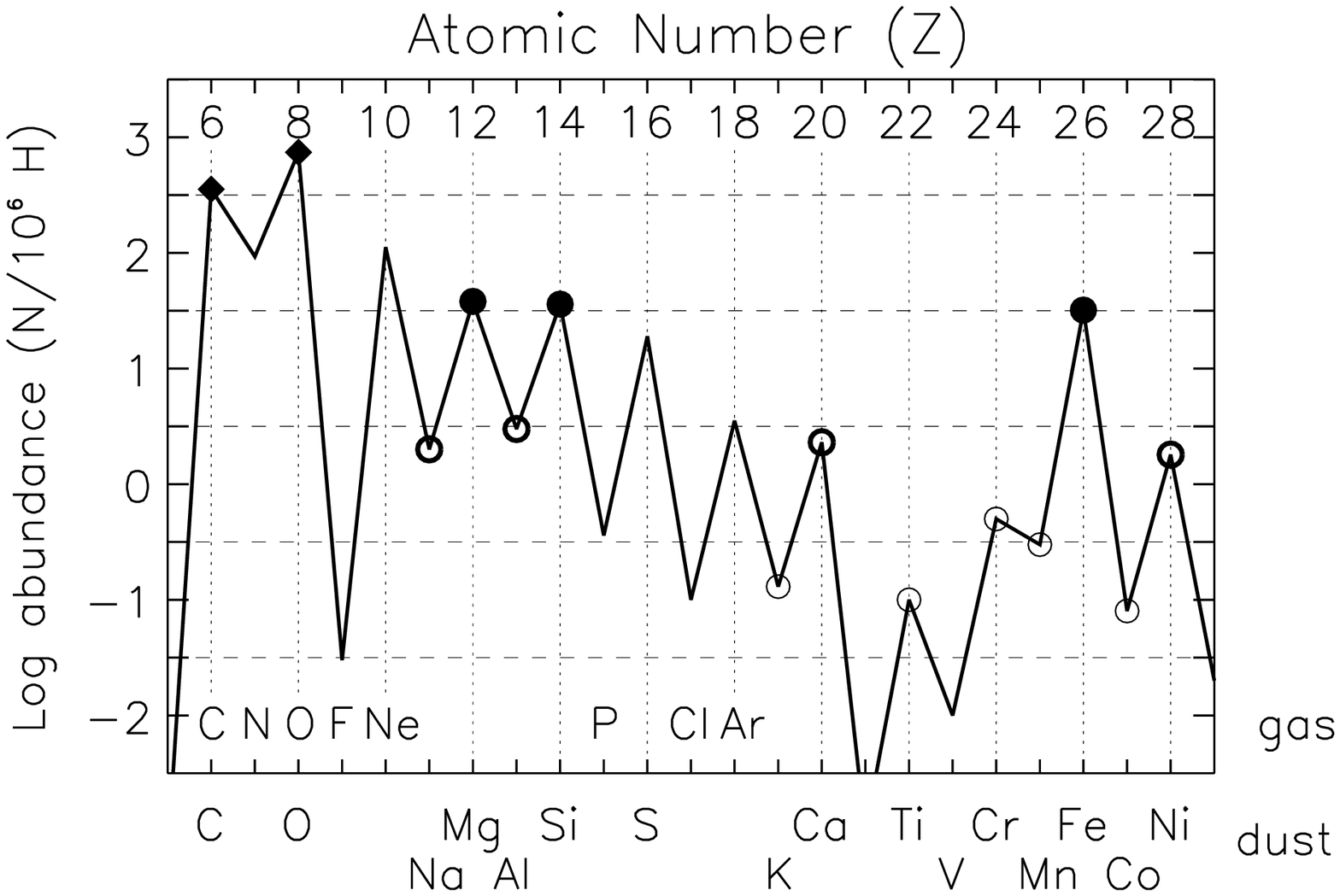}}
  \caption{Solar abundances in parts per million (ppm $\rm \equiv
  N/10^6 \, H$) {\it vs.}  atomic number [Anders \& Grevesse 1989;
  with C, N and O data from Grevesse and Noels 1993], ({\it solid
  line}). The elements labelled above (below) the x-axis indicate
  those elements with a `preference' for the gas (solid) phase. The
  `sawtooth' pattern, with peaks at even Z values, arises from the
  nucleosynthetic formation of the elements from He nuclei,
  $^4_2He^{++}$. The dust-forming elements can be grouped into primary
  (C and O with abundances $\geq$ 300 ppm, {\it filled diamonds}),
  major (Mg, Si and Fe with abundances $\sim$ 30 ppm, {\it filled
  circles}), minor (Na, Al, Ca and Ni with abundances $\sim$ 3 ppm,
  {\it heavy circles}), and trace (K, Ti, Cr, Mn and Co with
  abundances $\sim 0.1 - 0.3$ ppm, {\it light circles}) dust
  constituents.}
\end{figure}

\newpage
\begin{figure}
  \centerline{\epsfxsize=14.0cm \epsffile{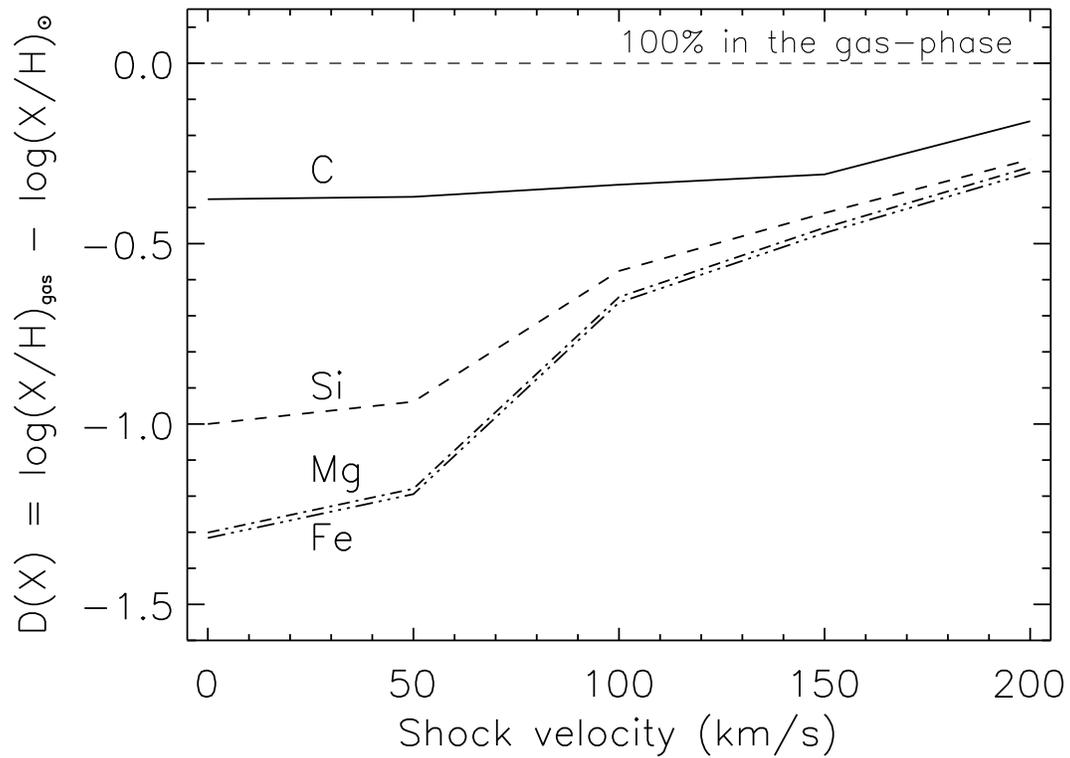}}
  \caption{Depletions of the dust-forming elements as a function of
  shock velocity, assuming the results of Jones, Tielens \& Hollenbach
  [1996]. Carbon ({\it solid}), silicon ({\it short-dashed}),
  magnesium ({\it dotted-dashed}), and iron ({\it triple
  dotted-dashed}).  The preshock fractions of the elements assumed to
  be in dust are O (0.16), C (0.58), Fe (0.95), Si (0.90), and Mg
  (0.95) [Draine and Lee 1984]. The results expressed in this form are
  independent of the assumed reference abundances. The sputtering of
  the silicate elements was assumed to be in their stoichiometric
  ratios.}
\end{figure}

\newpage
\begin{figure}
  \centerline{\epsfxsize=14.0cm \epsffile{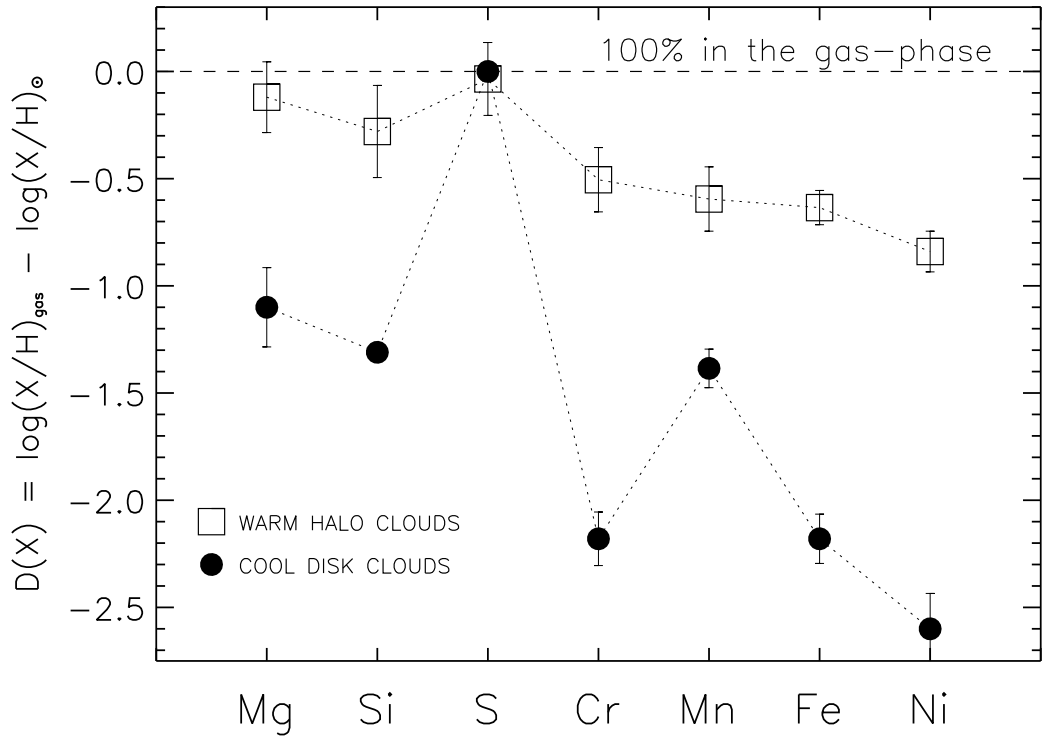}}
  \caption{Gas phase abundances for the major (Mg, Si and Fe), minor
  (Ni) and trace (Cr and Mn) dust-forming elements and S for diffuse
  lines of sight lines through cool clouds in the Galactic disk ({\it
  filled circles}) and warm clouds in the Galactic halo ({\it open
  squares}) for Solar reference abundances.  The vertical bars
  indicate the ranges of the observed values; data taken from Savage
  and Sembach [1996]. For the Mg data we have adopted the current best
  estimate for the Mg$^+$ oscillator strength from Fitzpatrick
  [1997].}
\end{figure}

\newpage
\begin{figure}
  \centerline{\epsfxsize=14.0cm \epsffile{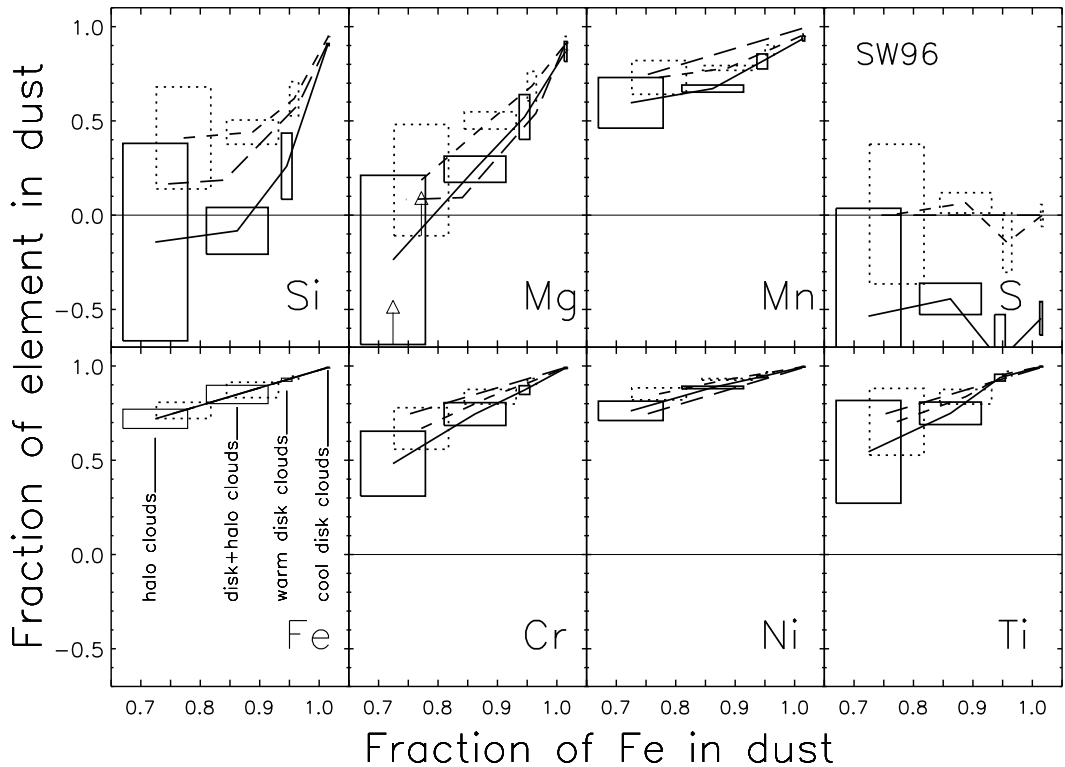}}
  \caption{Fraction of an element in dust normalized to that of Fe for
  the major (Mg, Si and Fe), minor (Ni) and trace (Ti, Cr and Mn)
  dust-forming elements and sulfur. The Snow and Witt [1996] reference
  abundances (SW96) were used, with the adoption of 2/3 of the solar
  abundance for Mn. Data is presented for the four types of diffuse
  cloud defined by Savage and Sembach [1996], namely; cool disk, warm
  disk, disk+halo and halo. The boxes indicate the ranges of values
  over a number of lines of sight, and the lines connect the mean
  values for each cloud type. All data is taken from Savage and
  Sembach [1996], but the Mg abundances have been updated using the
  Mg$^+$ oscillator strength of Fitzpatrick [1997]. The dotted boxes
  and short-dashed lines indicate the same data but assume solar
  reference abundances. The long-dashed lines show the results of a
  model core/mantle particle fit to the data assuming SW96 abundances
  with maximum dust phase abundances of 0\% for S, 90\% for Mg and Si,
  and 99\% for Fe and all other elements. We have assumed that 80\% of
  Mg and 70\% of Si are in (Mg$_{0.9}$,Fe$_{0.1}$)$_2$SiO$_4$ mantles
  and that the core consists of Fe-rich olivine and oxides. The Mn,
  Cr, Ni and Ti fractions in each component are the same as for Fe.
  In the model the sputtering erosion of the elements was assumed to
  be in their stoichiometric ratios.}
\end{figure}

\setcounter{figure}{0} 


\begin{planotable}{cccccc}
  \tablewidth{20pc}
  \tablecaption{Reference abundances for a selection of elements,
  including all the major dust-forming elements, presented as the
  number of atoms, N, per million H atoms ( N/10$^6$ H $\equiv$ ppm
  ). Z is the atomic number and A is the atomic mass of the element.}
  \tablehead{
  \multicolumn{1}{l}{ } & 
  \multicolumn{1}{c}{ } & 
  \multicolumn{1}{c}{ } &
  \multicolumn{3}{c}{Abundance (ppm)} \\[.3ex]
   \cline{4-6}\\[-1.6ex]
  \multicolumn{1}{c}{Element} &
  \multicolumn{1}{c}{Z} &
  \multicolumn{1}{c}{A} &
  \multicolumn{1}{c}{Solar  \tablenotemark{a}} &
  \multicolumn{1}{c}{B star \tablenotemark{b}} &
  \multicolumn{1}{c}{SW96   \tablenotemark{c}} }
\tablenotetext{a}{Anders \& Grevesse [1989]; except for C, N and O which
 are from Grevesse \& Noels [1993].}
\tablenotetext{b}{Field B star abundances from Snow \& Witt [1996] and 
 references therein.}
\tablenotetext{c}{Snow \& Witt [1996].} 
\startdata
  C  &  6 & 12.0 & 355   & 204    & 214     \nl
  N  &  7 & 14.0 &  93   &  68    &  66     \nl
  O  &  8 & 16.0 & 741   & 380    & 457     \nl
  Mg & 12 & 24.3 &  38   &  23.4  &  25     \nl
  Si & 14 & 28.1 &  36   &  15.8  &  18.6   \nl
  S  & 16 & 32.1 &  19   &  11.7  &  12.3   \nl
  Ca & 20 & 40.1 &   2.3 &   1.70 &   1.58  \nl
  Ti & 22 & 47.9 &   0.1 &   0.06 &   0.065 \nl
  Cr & 24 & 52.0 &   0.5 &   0.32 &   0.32  \nl
  Mn & 25 & 54.9 &   0.3 &   ---  &   ---   \nl
  Fe & 26 & 55.8 &  32   &  30.9  &  26.9   \nl
  Ni & 28 & 58.7 &   1.8 &   1.15 &   1.12  
\end{planotable} 

\begin{planotable}{lcccccc}
  \tablewidth{40pc}
 \tablecaption{The elemental ratios for IS silicate/oxide grain components.
  \tablenotemark{a}}
  \tablehead{
  \multicolumn{1}{l}{ }                &
  \multicolumn{2}{c}{Solar reference \tablenotemark{b}} &
  \multicolumn{2}{c}{B star reference\tablenotemark{c}} &
  \multicolumn{2}{c}{SW96 reference  \tablenotemark{d}} \\[.3ex]
   \cline{2-3} \cline{4-5} \cline{6-7} \\[-1.6ex]
  \multicolumn{1}{l}{Component}        &
  \multicolumn{1}{c}{(Mg+Fe)/Si}       &
  \multicolumn{1}{c}{Mg/Fe}            &
  \multicolumn{1}{c}{(Mg+Fe)/Si}       &
  \multicolumn{1}{c}{Mg/Fe}            &
  \multicolumn{1}{c}{(Mg+Fe)/Si}       &
  \multicolumn{1}{c}{Mg/Fe}            }
\tablenotetext{a}{Data from Savage and Sembach [1996] with Mg
 abundances updated using the Mg$^+$ oscillator strengths from
 Fitzpatrick [1997].}
\tablenotetext{b}{Anders \& Grevesse [1989].}
\tablenotetext{c}{Field B star abundances from Snow \& Witt [1996] and 
 references therein.}
\tablenotetext{d}{Snow \& Witt [1996].} 
\startdata
  Overall          &  2.0  &  1.1  & 3.5  &  0.6  &  2.9     &  0.8  \nl 
  Most refractory  &  2.1  &  0.3  & 4.7  &   0   & $\infty$ &   0   \nl 
  Least refractory &  1.8  &  4.1  & 1.8  &  4.1  &  1.8     &  4.1
\end{planotable} 

\begin{planotable}{lcc}
 \tablewidth{40pc}
 \tablecaption{H$^+$ implantation parameters for the gas behind a
  100~km/s shock wave and for cosmic rays.}
  \tablehead{
  \multicolumn{1}{l}{Parameter}      &
  \multicolumn{1}{l}{Shock wave  \tablenotemark{a}} &
  \multicolumn{1}{l}{Cosmic rays \tablenotemark{b}} }
\tablenotetext{a}{Data from Jones, Tielens and Hollenbach [1996].}
\tablenotetext{b}{Data from Moore [1999].} 
\startdata 
  Abundance (fraction)    & $\sim$ 0.9   & $\sim$ 0.9      \nl
  Energy                  & $\sim$ 50 eV & $\sim$ 1 MeV    \nl
  Implantation range      & $\sim$ 30\AA & $\sim$ 1 $\mu$m \nl
  Impacts/atom/10$^8$ yr  & $\sim$ 10    & $\sim$ 0.1      \nl
  Type of process         & stochastic   & continuous     
\end{planotable} 

\begin{planotable}{cccccc}
 \tablewidth{40pc}
 \tablecaption{The dust composition in the local ISM (data taken from
   Frisch et al. [1999]).}
  \tablehead{
  \multicolumn{1}{l}{ } &
  \multicolumn{5}{c}{Abundances (ppm) in a given phase \tablenotemark{a}}
   \\[.3ex]
   \cline{2-6}\\[-1.6ex]
   Element/Ratio & Total & Gas phase & Core+Mantle & Core & Mantle }
\tablenotetext{a}{Assuming the Snow and Witt [1996] reference abundances.} 
\startdata
   Mg        & 25  &  7  & 18  & 12  & 6   \nl
   Si        & 19  & 12  &  7  &  3  & 4   \nl
   Fe        & 27  &  3  & 24  & 13  & 11  \\[0.3cm]
  (Mg+Fe)/Si & 2.7 & 0.8 & 6.0 & 8.3 & 4.3 \nl
   Mg/Fe     & 0.9 & 0.8 & 0.9 & 0.5 & 2.3 
\end{planotable}

\clearpage

\end{document}